\documentclass[aps,preprint,showpacs,a4paper]{revtex4}

\usepackage{epsfig}
\usepackage{subfigure}

\newcommand{\bdm}{\begin{displaymath}}
\newcommand{\edm}{\end{displaymath}}
\newcommand{\benl}{\begin{equation}}
\newcommand{\be}[1]{\begin{equation}\label{#1}}
\newcommand{\ee}{\end{equation}}
\newcommand{\bea}{\begin{eqnarray}}
\newcommand{\eea}{\end{eqnarray}}
\newcommand{\ba}{\begin{array}}
\newcommand{\ea}{\end{array}}

\begin{document}

\title{Adsorption  of Indium on a  InAs wetting layer deposited on the
  GaAs(001) surface}

\author{Marcello Rosini}
\email{mrosini@unimo.it}
\affiliation{Dipartimento di  Fisica  dell'universit\'a degli studi di
  Modena e Reggio Emilia and S3 research center of CNR-INFM, via Campi
  213/A, 41100 Modena, Italy}

\author{Rita Magri}
\affiliation{Dipartimento di  Fisica  dell'universit\'a degli studi di
  Modena e Reggio Emilia and S3 research center of CNR-INFM, via Campi
  213/A, 41100 Modena, Italy}

\author{Peter Kratzer}
\affiliation{Fachbereich Physik, Universit\"at Duisburg-Essen, Germany}

\begin{abstract}
  In this  work we perform a first-principles  study of the adsorption
  properties of an In adatom deposited on $1.75$ monolayers (ML) InAs,
  forming  a wetting  layer   on   GaAs$(001)$  with the     $\alpha_2
  (2\times4)$ or $\beta_2  (2\times4)$ reconstruction.  The structural
  properties of these reconstructions have  been studied: we determine
  the equilibrium  geometry  of the surfaces  and  their stability for
  various  growth conditions.  We   have then carried  out a  detailed
  study of the  potential  energy surface (PES)  for  an In adsorbate,
  finding the minima and the  saddle points.  The main characteristics
  of the  PES and the bonding configurations  of the In  adatom on the
  surface are analyzed by comparing with analogous studies reported in
  the literature, trying  to   extract the  effects  due to:  (i)  the
  compressive strain to which the InAs adlayer  is subjected, (ii) the
  particular  surface reconstruction,    and (iii) the  wetting  layer
  composition.  We found that, in general, stable adsorption sites are
  located  at: (i) locations  besides the As in-dimers, (ii) positions
  bridging two As   in-dimers,  (iii) between two  adjacent  ad-dimers
  (only in $\beta_2$), and  (iv) locations bridging two  As ad-dimers.
  We   find  also other  shallower   adsorption  sites  which are more
  reconstruction specific due to  the lower symmetry of the $\alpha_2$
  reconstruction compared to the $\beta_2$ reconstruction.
\end{abstract}


\date{\today}

\maketitle


\section{Introduction}
Thanks to the strong charge carrier  confinement, QDs have atomic-like
properties    that could be  useful  for  applications  in optical and
optoelectronic  devices, quantum  computing, and  information storage.
One of the most challenging problems for quantum dot formation, is the
control of their  shape, composition and density.   From this point of
view the  abundant   experimental knowledge has  not  yet   produced a
sufficiently deep understanding of the physics of formation of quantum
dots that could guide the way to  control the growth process.  In this
situation first-principles simulations can be helpful in sheding light
on the atomistic mechanisms that lead to dot nucleation at surfaces.

In the present paper, we address one of the most-studied systems: InAs
quantum   dots grown  on  a  GaAs(001)   substrate, where the  lattice
mismatch between InAs  and GaAs  is  of  the order  of $7$\%.   Strain
relaxation at the surface acts as a driving  force for a self-assembly
process, the so  called Stranski-Krastanov  growth mode \cite{stran},
with  the deposition of an  atomically thin InGaAs  wetting layer (WL)
and subsequent surface mass transport accompanied by a 2D to 3D growth
transition.

We focus our attention on the structural properties of the WL on which
quantum  dots  nucleate, and on  the  interaction between a  single In
adatom and the surface, in order  to find where the diffusion barriers
are located and what the possible adsorption sites on the surface are,
as a function of the wetting layer reconstruction.

We   have  chosen to  investigate mainly   the  $\alpha_2 (2\times 4)$
reconstruction of  the $(001)$ surface because  of its stability under
In-rich  conditions  and under compressive strain  \cite{ratsch, add1,
  add2} which is the case for the InAs WL  grown on GaAs.  Moreover, a
$(2\times  4)$  reconstruction was  found \cite{balzarotti, grandjean,
  add3,  add4} to occur on  the $(001)$ wetting  layer at the onset of
dot nucleation in a  In-rich WL condition.  We  have modeled a WL at a
InAs  coverage  $\theta=1.75$ monolayer  (ML)  which  is close  to the
critical value  corresponding to the  2D to 3D transition \cite{2to3}.
The structural and  geometrical properties of  this  surface have been
investigated.  Using the  $\alpha_2$ reconstruction we have calculated
the PES for a single  In adatom, with  the aim to understand what  are
the implications of the WL morphology  (strain and composition) on the
adsorption  and surface mobility  of the deposited  In.  To this end a
comprehensive comparison   with results   reported previously in   the
literature for Ga and In  adatoms and other surface reconstructions is
given.  In this  context, a detailed  analysis of the differences  and
analogies between the $\alpha_2$  and the $\beta_2$ reconstructions is
pursued.

In the   literature similar calculations have  been  reported for GaAs
$[100]$ homoepitaxy \cite{kratzer}.   The issue of the  WL composition
for  In  adsorption  has   been  addressed  only by  Penev    et.  al.
\cite{penevtesi, penevpap},  but for  a  much lower In  coverage of
$\theta=0.66$ ML.

The  paper  is  organized as  follows:  in  section  II we detail  our
theoretical  approach. Section III  reports  the obtained results for:
(i) the WL surface reconstruction and energy; (ii) the PES of a single
In adatom on the $\alpha_2$  and the $\beta_2$ reconstructed WL; (iii)
the determination  of the barrier  positions and heights  between each
couple of    adsorption sites; (iv) a  detailed    study of the atomic
bonding of the  In  adatom to the  WL. In  section IV our  results are
compared with those  regarding other surface reconstructions  in order
to extract  a   rationale for the    dependence on In adsorption   and
diffusion on the local structure  of  the $(001)$ surface. Finally,  a
summary is given at the end of the paper.

\section{Theoretical method}


We performed  first principles calculations within  Density Functional
Theory in the local density  approximation (DFT-LDA) \cite{lda}  using
the exchange   and correlation   potential   of  Ceperley and    Alder
\cite{cepa} as parametrized by Perdew and  Zunger \cite{pz}.  For this
kind of  calculations  both LDA  \cite{ratsch, penevc4x4,   bechstedt,
  wang, lepage}  and GGA  \cite{ruggerone, giappi1, giappi2, penevpap,
  kratzer} have been used in the literature.  It has been demonstrated
\cite{add5, add7}  that   both  methods give  qualitatively   the same
picture  for  the  description of adatom   adsorption   on a  surface,
although LDA usually gives larger values for binding energies than GGA
\cite{add6}.    Only the details    can  be different:  LDA  tends  to
overestimate  the binding energy but   for lattice constants,  elastic
moduli and surface energies the   LDA calculated values are  generally
closer   to  the  experimentally    determined ones.    We have   used
norm-conserving pseudopotentials   treating   the   outermost  s-  and
p-shells of Ga, In  and As  as  valence electrons, and the  electronic
wave  functions were expanded in  plane-waves,  with a $18$ Ry kinetic
energy cutoff.  The energy  cutoff has been tested  in order to  reach
convergence for  the lattice bulk  properties  of GaAs and  InAs.  The
core corrected atomic pseudopotentials have been tested on bulk Ga, In
and As,  where  the determined equilibrium  configurations and elastic
moduli compare well with the experimental data.

The equilibrium lattice parameters obtained for the GaAs and InAs bulk
phases are  $a_0=5.609$ \AA\ and $a_0=5.906$  \AA,  which are slightly
smaller  than  the  experimental  ones. Thus   our  calculated lattice
mismatch is $5.4$\%,   smaller than the  experimental value.   This is
going to underestimate the WL  strain. In sec.  IV we will compare our
results  with results \cite{penevc4x4,   penevpap, kratzer} where this
effect was instead overestimated.

Starting from the GaAs structure, we have set  up the $(001)$ oriented
supercell containing 4 layers  of GaAs,  covered  with 1.75 layers  of
InAs  arranged  according to   the   $\alpha_2 (2\times   4)$  surface
reconstruction.  The lower layer of Ga  atoms is kept fixed during the
cell  relaxation,  in order   to  mimic  the   constraint due  to  the
underlying  semi-infinite   bulk,  and it is    passivated with pseudo
hydrogen atoms of $1.25$ electron charge.  The  slab is repeated along
the $(001)$ direction with a  periodicity $5 a_0$  and a separation of
about 10 \AA\ of vacuum.   As for the number of  GaAs layers, we  have
found that only two layers suffice to  obtain a correct description of
the PES, minima  and maxima positions.   The refinement using two more
GaAs layers produces a difference  in the  barrier heights all  within
$40$ meV.  Brillouin-zone (BZ) integration was carried out using a set
of special $k$-points  equivalent to  16  points  in the $1\times   1$
surface BZ. A smearing of $0.02$ Ry has  been used in order to account
for the possible metalization of the surface electronic structure.

\begin{figure}[t]
\centering
\includegraphics*[width=5.5cm]{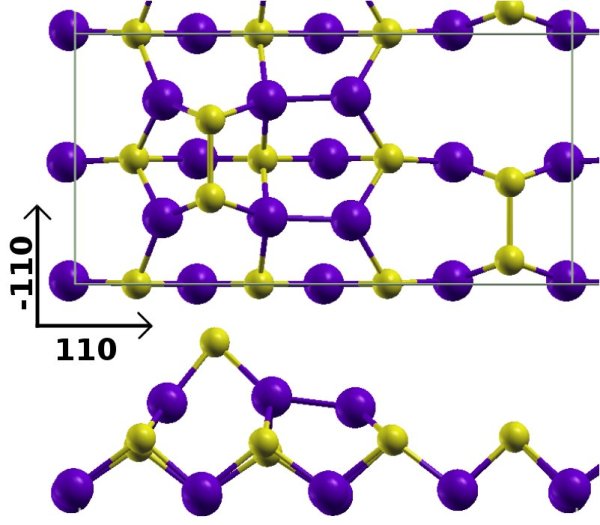}
\hskip5mm
\includegraphics*[width=5.5cm]{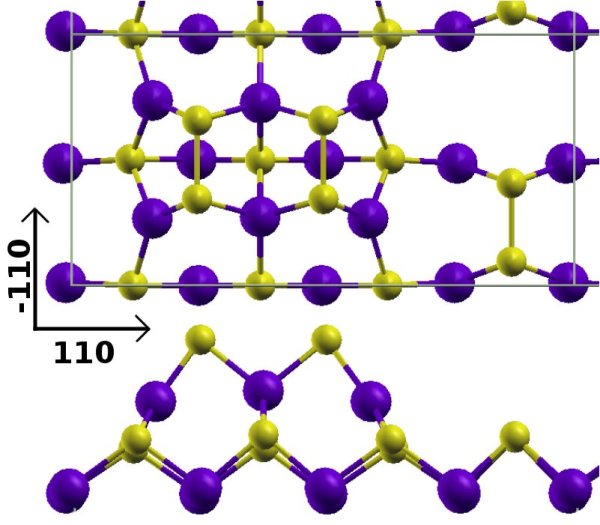}
\caption{(Color online)  Representation of the  $\alpha_2 (2\times 4)$
  (left) and  $\beta_2  (2\times 4)$ (right)  surface reconstructions.
  The shown geometries are those optimized. It  is evident the bending
  of the  the  In-In  bond   in the  top    layer in  the   $\alpha_2$
  reconstruction. Dark balls and light balls represent  In and As ions
  respectively.}
\label{struc}
\end{figure}

In the $\alpha_2(2\times 4)$  reconstruction the last  complete atomic
layer is constituted   by As atoms.  A $\theta=0.75$  cation  layer is
deposited over it. This layer is then terminated with one As dimer (As
ad-dimer) on top  (see  Fig.  \ref{struc}).   The uncovered As  on the
last complete  layer dimerizes along the  $[\bar 1 1 0]$ direction (As
in-dimer).  The  $\beta_2(2\times 4)$  reconstruction is  similar  but
presents one  additional As ad-dimer  bridging  the two  remaining In
rows on  the top of the   surface (see Fig.   \ref{struc}).   It has a
larger  symmetry  than   the $\alpha_2$  reconstruction,  being mirror
symmetric   with respect to the  $(110)$  plane passing through the As
in-dimers.  All the examined structures have been  relaxed in order to
find the  equilibrium geometries, until  all the forces  acting on the
atoms were less than $2.5\cdot10^{-3}$ eV/\AA.

The  PES  of an  In adatom  is calculated by  relaxing  the adatom $z$
component together with  all the surface  degrees of freedom while the
in-plane $x,\ y$ coordinates are kept fixed. We  have set up a grid of
$8\times4$ points,  corresponding  to a  step of   $2$ \AA, along  the
$(110)$ and $(\bar110)$ directions respectively, and for each point we
have  found the minimum  energy  configuration for the adsorbate  plus
surface.    Then we have  interpolated the  grid  data with a bi-cubic
spline algorithm in order to find the  positions of the minima and the
saddle  points.  The  exact   values  of the minima   have  then  been
determined by further relaxing all the three coordinates of the adatom
together  with the surface.   To minimize the adsorbate interaction we
have doubled the surface unit cell along the $(\bar110)$ direction.


Since one of  the objectives of this   work is ultimately  to study In
diffusion on the WL  within  the  framework  of the Transition   State
Theory \cite{tst1, tst2, tst3}, the minimum energy paths (MEP) and the
height of each energy barrier between every couple of adsorption sites
have to be  calculated, leading to  a more  detailed  knowledge of the
PES.  This task has been accomplished by using the Nudged Elastic Band
method  (NEB).    This method is  able   to find the  MEP  between two
adsorption  sites, by simulating a string  of replicas  of the system,
where the  different images  are  one linked  to the other  by springs
\cite{neb, neb2}.  By minimizing the energy associated to the path, an
accurate description  of the MEP   and thus  of   the saddle point  is
obtained.


All   the ab-initio  calculations  were  performed using the  ESPRESSO
simulation package \cite{webespresso}.

\section{Results}

\subsection{Structure of $(001) (2 \times 4)$ surface reconstructions}
\begin{figure}[t]
\centering
\includegraphics*[width=8cm]{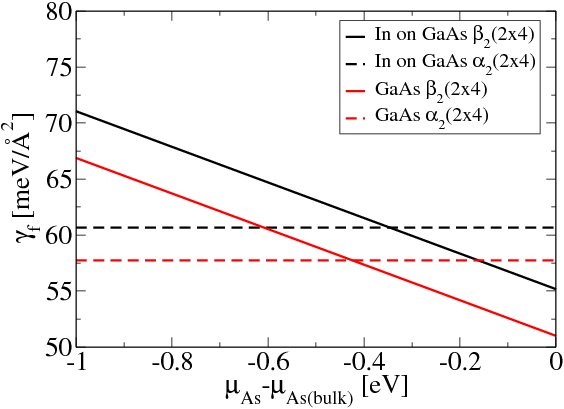}
\caption{(Color    online)   Surface energy  phase    diagram  for the
  $\alpha_2(2\times 4)$  and  $\beta_2(2\times 4)$ reconstructions  of
  InAs/GaAs$(001)$ and GaAs$(001)$.}
\label{phase}
\end{figure}

We have calculated the optimized geometries and the formation energies
of the $\alpha_2(2\times4)$  and $\beta_2(2\times  4)$ reconstructions
of the bare GaAs  surface and of GaAs  covered with a $\theta=1.75$ ML
InAs  layer.   The surface energies   vs.  the As  chemical  potential
$\mu_{As}$ (whose  value is related to the  growth conditions  such as
growth  temperature, cation/anion flux  ratios) are calculated through
the expression \cite{fff}
\begin{equation}
\gamma_{\mathrm f}=
\frac 1 A \left( E_\mathrm{tot}-
N_{\mathrm{Ga}}\mu_{\mathrm{GaAs}}
-N_{\mathrm{In}}\mu_{\mathrm{InAs}} \right)-\mu_{\mathrm{As}}
\frac{N_{\mathrm{As}}-N_{\mathrm{Ga}}-N_{\mathrm{In}}}{A}\ ,
\end{equation}
($\mu_{\mathrm{InAs}}=0$,   $N_{\mathrm{In}}=0$   for pure GaAs).  The
obtained   surface energies are  reported  in Fig.  \ref{phase}.  Here
$E_\mathrm{tot}$ is the total energy of the slab, $\mu_\mathrm{xx}$ is
the  chemical   potential of    the    bulk phase  of  material    xx,
$N_\mathrm{y}$ is  the number of atoms  of element y in  the supercell
and $A$ is  the  surface unit cell  area.   A set of   calculations is
performed for  a slab with  two hydrogen passivated surfaces, in order
to  subtract   the  contribution  of  the   bottom  surfaces   of  the
reconstructed slabs to be  investigated, hence $E_\mathrm{tot}$ is the
net contribution  of the reconstructed slab.  Elemental  As, Ga and In
are  calculated using their  ground state structures: respectively the
rhombohedral A$7$ structure for As, $\alpha$-Ga and bct In.  All these
calculations have been done using high  convergence standards.  We see
that the range  of  stability of  the $\alpha_2$ phase,  when  GaAs is
covered with the InAs ad-layer,  is larger than  that of the bare GaAs
surface.  This means  that the formation  of the $\alpha_2$ phase will
be favored over the  $\beta_2$ phase over a larger   range of As  flux
values (or   surface growth conditions), when  the  InAs WL is formed.
This  is    in   agreement with  previous     calculations   showing a
stabilization  of  the   $\alpha_2$ phase for    InAs under  isotropic
compressive strain  \cite{ratsch,penevtesi} which is  the condition of
the InAs layer grown on GaAs.

Regarding the  atomic  structures, the   comparison of  the  distances
between the atomic  planes, in the InAs covered  GaAs case, shows that
the values are systematically larger than  those obtained for the pure
GaAs surface by  about $10$ \%, owing  to the stress originated by the
InAs/GaAs lattice mismatch.  In  particular, we remark the buckling of
the bond between  the In atoms in  the uncompleted In  layer found for
the $\alpha_2$   reconstruction  (see Fig.  \ref{struc})   due  to the
missing  (with respect  to  the $\beta_2$   reconstruction)  second As
ad-dimer. It  was shown  (\cite{ratsch})  that this  buckling plays a
role in the stabilization of the $\alpha_2$ phase.

\subsection{The potential energy surface of the single In adatom}
In the following we describe our  calculations for both the $\alpha_2$
and $\beta_2$ surface reconstructions,  with an accurate description of
the adsorption sites and hopping barriers.

\subsubsection{$\alpha_2(2\times 4)$ surface reconstruction}
\begin{figure}[t]
\centering
\includegraphics*[width=10cm]{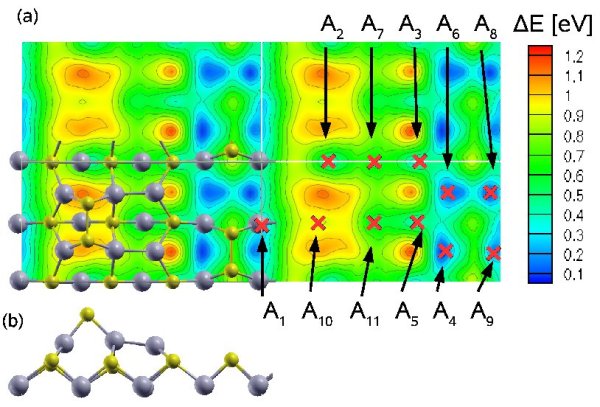}
\caption{(Color online)   (a) PES of  an In   adatom on the $\alpha_2\
  (2\times4)$ surface. The PES is drawn for $4$ surface unit cells and
  a  top view of the atomic  structure is also  shown.  The minima are
  labeled $A_i$. (b) Side view of the surface layer.}
\label{pes}
\end{figure}

\begin{figure}[t]
\centering
\subfigure[\label{ma}]
{\includegraphics*[width=5.5cm]{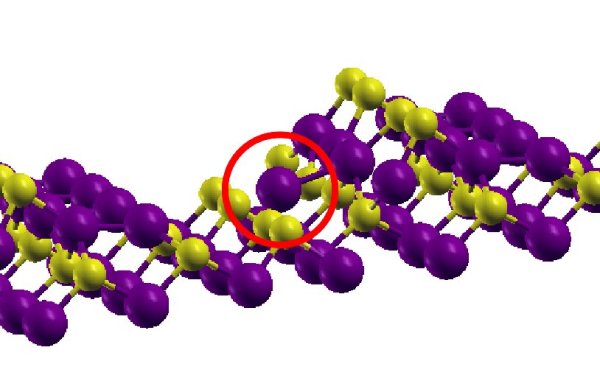}}\quad
\subfigure[\label{mc}]
{\includegraphics*[width=5.5cm]{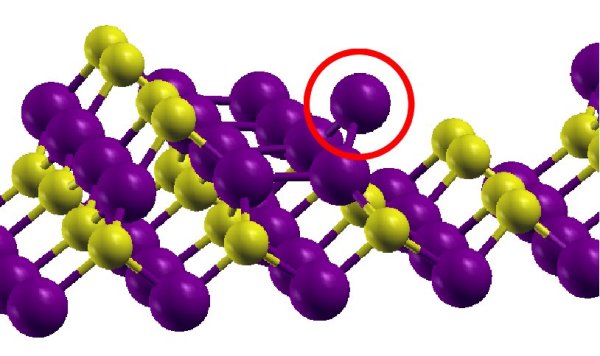}}
\subfigure[\label{mb}]
{\includegraphics*[width=5.5cm]{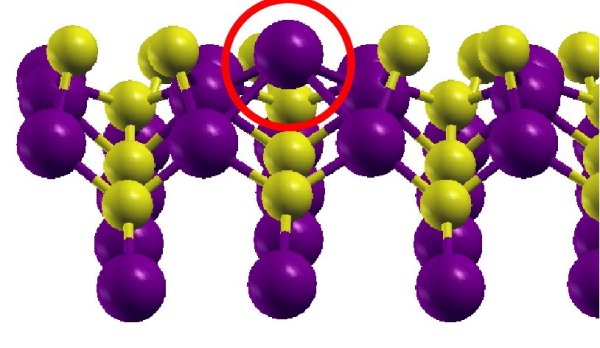}}\quad
\subfigure[\label{md}]
{\includegraphics*[width=5.5cm]{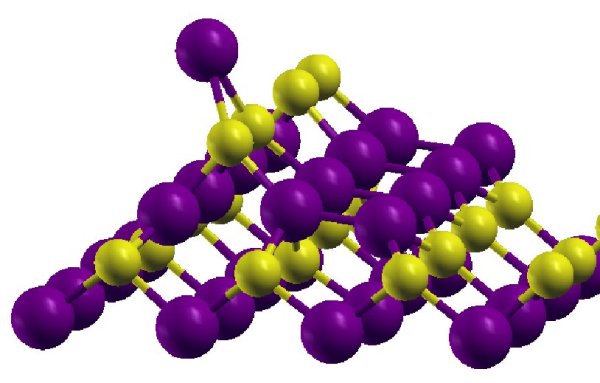}}
\caption{(Color online) Adsorption  configurations for the In adatom in
  different adsorption sites on the $\alpha_2$ surface reconstruction.
  (a)  Configuration  $A_8$ (b) Configuration $A_5$  (c) Configuration
  $A_2$ (d) Configuration $A_{10}$}
\label{a2-mins}
\end{figure}

For this surface reconstruction we have  found eleven adsorption sites
for the In adatom, and the related saddle points.  The energies of the
minima and  of the saddle points  are  given in Tables  \ref{tab1} and
\ref{tab2}, respectively, and a graphical representation of the PES is
reported in Fig.  \ref{pes}. In  the Tables, the energies are relative
to that of the deepest adsorption site A$_8$  put to $0$. Two main low
potential trenches are evident  along  the $[\bar110]$  direction,  at
both sides of the As in-dimer  row.  The deepest adsorption site A$_8$
is located in the trench near the In vacancy  of the surface layer and
is bonded to the two As atoms of the in-dimer  and to one In atom (see
Fig.   \ref{ma}).  Other  deep  minima (A$_4$,  A$_6$,  and A$_9$) are
located  along  the two deepest  trenches   and, together with  A$_8$,
represent  the  lowest    energy  set of     minima.  Their adsorption
configuration   is  similar to   that  of  A$_8$.    A second class of
adsorption sites (A$_1$,  A$_3$, A$_5$, A$_{11}$)  is characterized by
the In  adatom bonded to two other  In atoms (see the configuration at
minimum  $A_5$ in Fig.  \ref{mc}).  At  the minima A$_7$ and A$_2$ the
adatom is  bonded  to four In   atoms, reproducing the  bonding scheme
typical  of bulk Indium  (Fig.   \ref{mb}).  In these  last  cases the
corresponding surface electronic  structure is metallic.   To the last
class  of adsorption sites belongs  the shallow minimum A$_{10}$ where
the adatom  is  bonded  to the two    As  of the  ad-dimer  (see  Fig.
\ref{md}).  This configuration  plays the role  of a precursor for the
much stronger bonding of In into the ad-dimer (see below).

We have  accurately  checked that the shallow  minima  in  the PES are
actually minima and not saddle points of the energy landscape.

The location  of  the  calculated  barriers  T$_n$ is  shown  in  Fig.
\ref{pa}. The barriers are lower  along the $[\bar110]$ direction (the
highest barrier being $430$  meV and the lowest  being $52$ meV)  than
along  the orthogonal $[110]$ direction   (see Table \ref{tab2}). This
difference  is  mainly  ascribed to  the  presence of  the As ad-dimer
oriented  along the $[\bar110]$  direction,  that creates  an unstable
region  for the  In  adsorption, as can be   seen  in the  PES of  Fig
\ref{pes}.   From  these considerations  we  can preliminarily conclude
that adatom surface diffusion  should be strongly anisotropic with the
$[\bar110]$ direction along the trenches favored.

\subsubsection{$\beta_2(2\times 4)$ surface reconstruction}
Since  the symmetric part  of the $\beta_2$  has the same structure of
half  of the $\alpha_2$  surface,   the coordinates of the  adsorption
sites can be roughly estimated, and their  exact location can be found
with a further relaxation of the surface-adsorbate system.

\begin{figure}[t]
\centering
\subfigure[\label{pa}]{\includegraphics*[width=7cm]{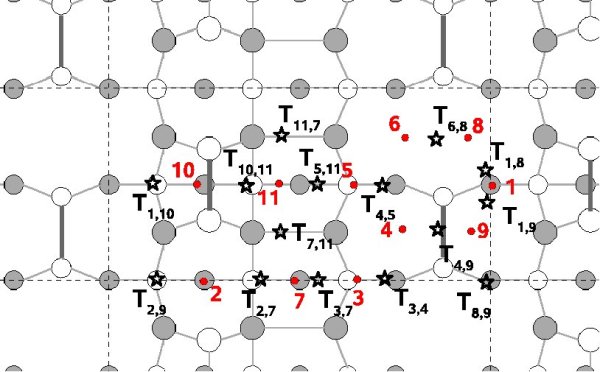}}
\qquad
\subfigure[\label{pb}]{\includegraphics*[width=7cm]{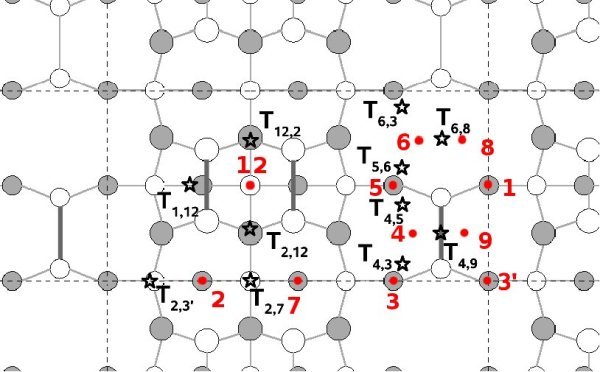}}
\caption{(Color online)  Schematic  view    of the adsorption    sites
  locations (only $i$ is written) on  the: (a) $\alpha_2 \ (2\times4)$
  and (b) $\beta_2  \ (2\times4)$ reconstructions.  The  saddle points
  are labeled as $T_{i,j}$.}
\label{mins-b2}
\end{figure}

Also for   the  $\beta_2$ surface reconstruction,  we   have  found 11
adsorption sites for the Indium  adatom (see Table \ref{tab3} and Fig.
\ref{pb}).  However, owing to  the symmetry properties of  the system,
those minima reduce to only 6 inequivalent minima.

\begin{figure}[t]
\centering
\subfigure[\label{ta}]{\includegraphics*[width=5.5cm]{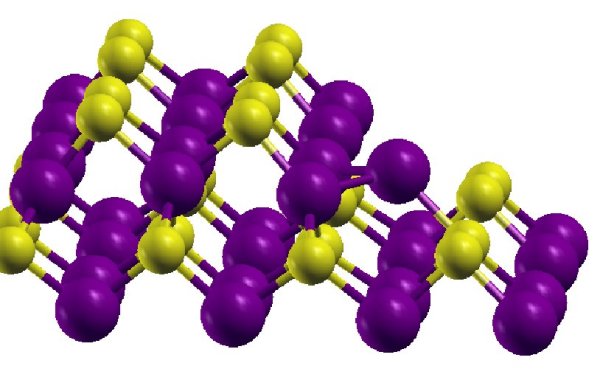}}
\subfigure[\label{tb}]{\includegraphics*[width=5.5cm]{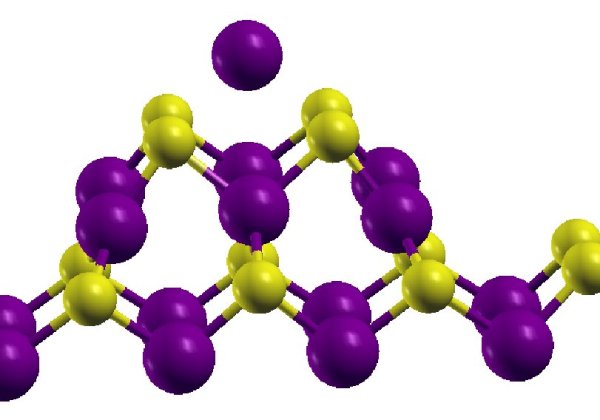}}
\caption{(Color online)  Adsorptions configurations for  the In adatom
  on  the $\beta_2$  surface  reconstruction.  (a) Configuration $A_5$
  (b) Configuration $A_{11}$}
\label{b2-mins}
\end{figure}

We first  notice that the  adsorption sites A$_{1}$, A$_{2}$, A$_{4}$,
A$_{6}$, A$_{7}$, A$_{8}$,   A$_{9}$, are in  strict correspondence to
those  of     the  $\alpha_2$    surface  reconstruction    (see   Fig
\ref{mins-b2}),  in  the region where the  two  structures are similar
i.e.  far from the  $\beta_2$ second ad-dimer.   The positions  of the
minima are only  slightly  affected by  the  presence of the  extra As
ad-dimer, and the  deepest adsorption  site  A$_{8}$ lies  in the same
position as for the $\alpha_2$ reconstruction.  The minima A$_{3}$ and
A$_{5}$   are  shifted   towards  the trench    with  respect  to  the
corresponding ones of the  $\alpha_2$ reconstruction.  This is due  to
the fact that the In adatoms in A$_{3}$ and  A$_{5}$ on the $\alpha_2$
reconstruction are bonded to  the buckled In at  the trench edge.  The
same In is not buckled in  $\beta_2$ due to  its bond to the second As
ad-dimer row  (compare Fig.  \ref{mc} to  Fig. \ref{ta}).  To quantify
the difference  between  the A$_1$-A$_9$  adsorption  sites in the two
reconstructions, we  give  in Table  \ref{tab5} the  adsorption energy
difference  $\Delta E=E^\beta(A_i)-E^\alpha(A_i)$  (negative  if In is
more  strongly    bonded in  the      $\beta_2$  than  in   $\alpha_2$
reconstruction)      and       the           displacement      $\Delta
r=\sqrt{\sum_i(r^\beta_i-r^\alpha_i)}$  between   the    corresponding
binding locations. The adsorption energy is defined as
\begin{equation}
E^\alpha(A_i)=E(\alpha_2+\mathrm{In}_{A_i})-E(\alpha_2)-E(\mathrm{In})\ .
\end{equation}
In   $\beta_2$,  A$_{3'}$ Fig \ref{mins-b2}  is   the symmetric of the
minimum A$_{3}$.  This  minimum  is  not  present in the    $\alpha_2$
reconstruction  because of the lower  symmetry. A new minimum A$_{12}$
is now present  for $\beta_2$ on top  of the two  ad-dimers and  it is
actually  quite stable, since the  In-adatom is fourfold bonded to the
surface  (see fig. \ref{tb}).  In this   case the confinement barriers
are calculated to be of the order of $135$ meV.  The configuration for
In in  the  adsorption sites  different from  those of  the $\alpha_2$
reconstruction, are given in Fig. \ref{b2-mins}.

Also in the $\beta_2$ case, two trenches in the  PES are evident along
the  $(\bar110)$  direction at  both sides  of  the  As in-dimer.  The
existence of the A$_{12}$ adsorption site on $\beta_2$ could originate
additional  diffusion  paths along the  same  direction connecting the
minimum  $A_{12}$  to    the  two  sites  $A_2$   and   $A_{7}$.   The
corresponding saddle point energies are  reported in Table \ref{tab4}.
We can  easily  see that the   barriers along $[\bar110]$, are  in the
range   $100$-$580$ meV and  are  lower than  the   barriers along the
$[110]$  direction  that   span  the  range  $100$-$760$  meV.    This
consideration  strongly indicates that  also for the $\beta_2$ surface
reconstruction, the In-adatom diffusion should be highly favored along
the $(\bar110)$ direction.

\subsection{Additional adsorption sites due to In insertion in the As dimers}
We know from the literature \cite{penevc4x4,ruggerone} that additional
minima could be  present, corresponding to  the In adatom breaking the
bond between the As  adatoms in a  dimer  and binding itself to  them.
This  process  requires overcoming an  energy  barrier,  thus its rate
should   depend  on the  growth  temperature.  These  sites  can be of
significant interest, since they introduce a local modification of the
PES that can  be very  important for  surface diffusion. For  example,
such sites could be in a  bridge location between two other adsorption
sites, thus introducing  a   further minimum in  an  otherwise thicker
barrier.  Such an additional adsorption site  can affect the diffusion
kinetics of the In adatom on the surface.

We  have considered   both   the As in-dimers  and   ad-dimers  of the
$\alpha_2$  and $\beta_2$ surface   reconstructions and determined the
energy necessary  to break  the As-As bond   and bind  the  In adatom.
First of all we have put In in the middle between the two As atoms, in
a vertical plane containing all the three atoms.  We have then relaxed
the system  obtaining the minimum  energy position  for the In adatom.
Finally we have performed different  NEB calculations in order to find
the energy barriers between these adsorption sites and the neighboring
sites.  We have  found that the  adsorption configuration with the  In
atom  bonded  to the two   arsenic of the  dimers are  in general more
stable than most of the  other adsorption sites (see Tables \ref{tab1}
and \ref{tab3}).

\subsubsection{$\alpha_2(2\times 4)$ surface reconstruction}
\begin{figure}[t]
\centering
\includegraphics*[width=7cm]{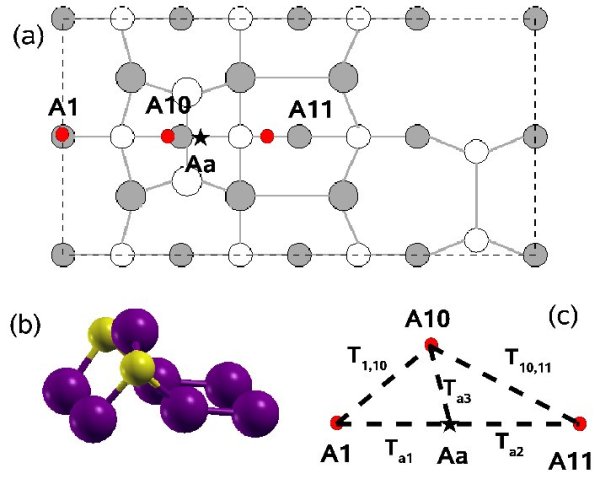}
\includegraphics*[width=7cm]{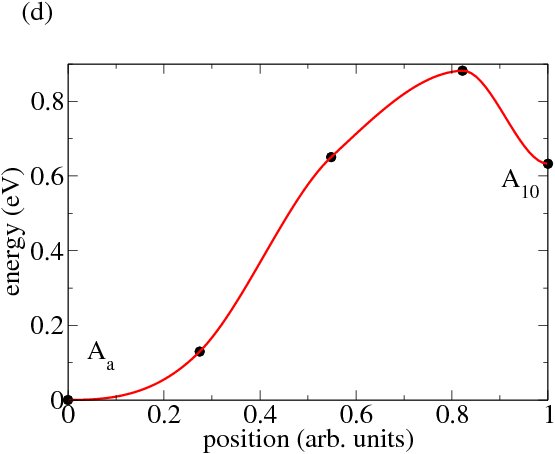}
\caption{(Color online) Additional adsorption site A$_a$ due to the In
  insertion into the As  ad-dimer  on the $\alpha_2$   reconstruction.
  (a) The location of the adsorption site is indicated by a star.  (b)
  Atomic configuration.  (c)  Scheme  of the transitions between   the
  A$_a$ adsorption site and the neighboring sites.  (d) Energy profile
  for barrier $T_{a3}$ from the precursor site A$_{10}$ to A$_a$.}
\label{a2-ad}
\end{figure}

\begin{figure}[t]
\centering
\includegraphics*[width=7cm]{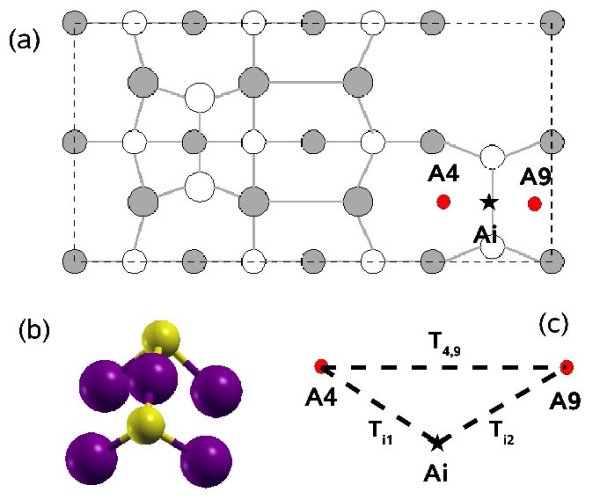}
\includegraphics*[width=7cm]{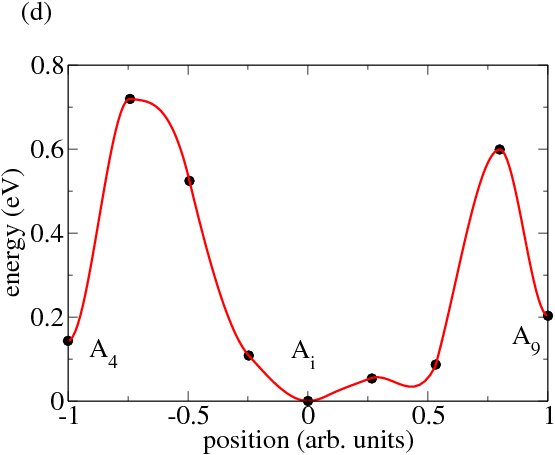}
\caption{(Color online) Additional adsorption site A$_i$ due to the In
  insertion  into  the   As   in-dimer on   the   $\alpha_2$   surface
  reconstruction. (a) The location of the adsorption site is indicated
  by a star. (c) Atomic configuration.  (b)  Scheme of the transitions
  between the A$_i$  adsorption site and the  neighboring sites.   (d)
  Energy  profile  for  barriers $T_{i1}$  and   $T_{i2}$ from the two
  besides in-dimer adsorption sites A$_4$ and A$_9$ to A$_i$.}
\label{a2-in}
\end{figure}

In this case  we  have found  2  additional  stable sites  for  the In
adsorbate.   The first  one   corresponds to the  breaking   of the As
ad-dimer and the second one to  the breaking of the  As in-dimer.  The
two  cases    are  shown in  figures   \ref{a2-ad}    and \ref{a2-in},
respectively.   Figures \ref{a2-ad}(a)  and  \ref{a2-in}(a)  show  the
location of these minima, indicated  as A$_a$ and A$_i$  respectively,
on the  surface  ($a=$ad-dimer $i=$in-dimer).  Figures  \ref{a2-ad}(b)
and \ref{a2-in}(b)  show  the atomic  configuration  of the  adsorbate
bonded to  the   As atoms in   the dimers.    The transitions  to/from
neighboring  adsorption sites are  also drawn   in the figures  (Figs.
\ref{a2-ad}(c)  and \ref{a2-in}(c))  and  the corresponding minima and
saddle  point energies are  given  in Tables \ref{tab1} and \ref{tab2}
respectively.  We can see that those minima are deep and stable, since
the   confining barriers  ($\simeq\    0.8$ eV),  pictured in  Figures
\ref{a2-ad}(d) and \ref{a2-in}(d) are high.  The energy barrier needed
to fall into A$_a$ from the site  A$_{10}$ (the precursor site) is not
too high (around $200$ meV)  and this process is  likely to occur even
at relatively low temperature.

\subsubsection{$\beta_2(2\times 4)$ surface reconstruction}
\begin{figure}[t]
\centering
\includegraphics*[width=7cm]{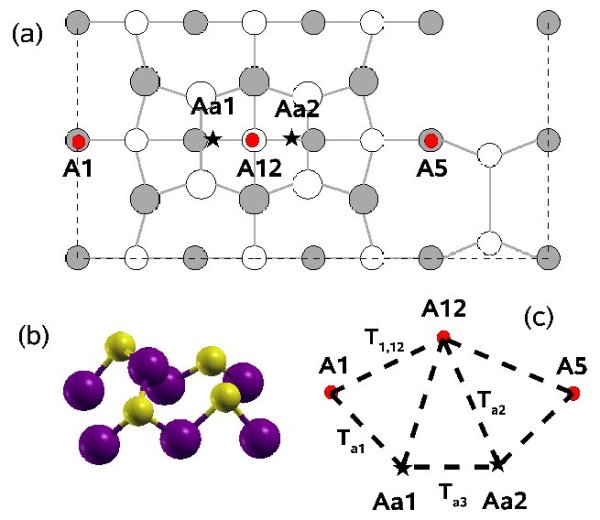}
\includegraphics*[width=7cm]{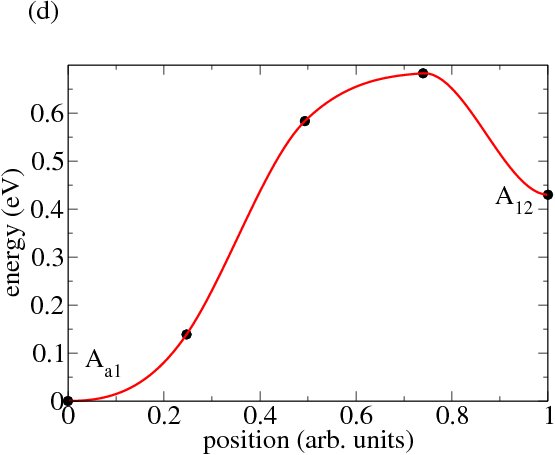}
\caption{(Color  online)  Additional   adsorption sites   A$_{a1}$ and
  A$_{a1}$ due  to  the  In  insertion  into the  As  ad-dimer  on the
  $\beta_2$ surface reconstruction. (a) The location of the adsorption
  sites are indicated by stars.  (b) Atomic configuration.  (c) Scheme
  of the transitions between the A$_{a1}$ and A$_{a1}$ adsorption site
  and the neighboring sites.  (d) Energy profile for barriers $T_{a2}$
  from the top ad-dimer adsorption site A$_{12}$ to A$_a$.}
\label{b2-ad}
\end{figure}

\begin{figure}[t]
\centering
\includegraphics*[width=7cm]{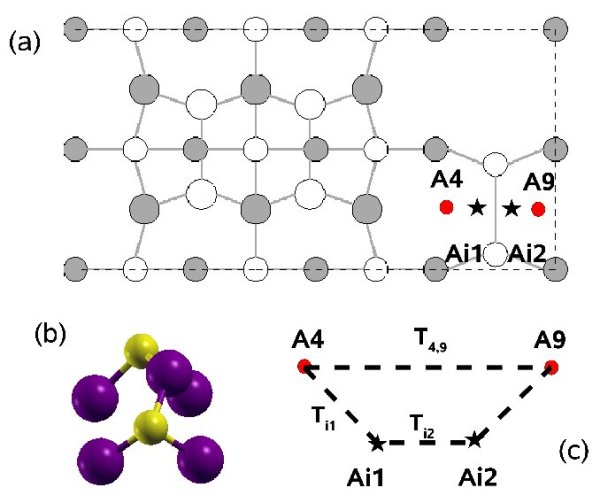}
\includegraphics*[width=7cm]{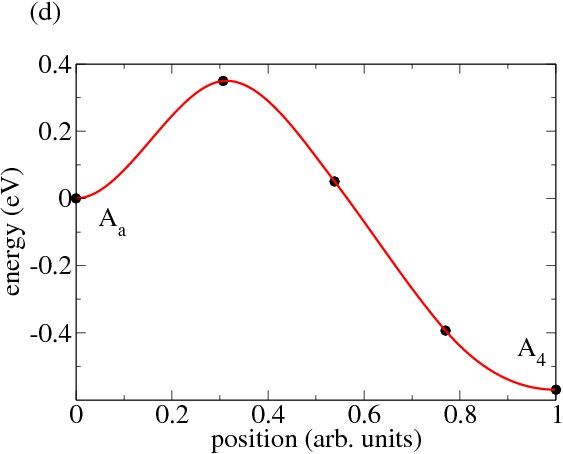}
\caption{(Color  online)   Additional  adsorption  sites  A$_{i1}$ and
  A$_{i2}$ due to  In insertion into the As  in-dimer on the $\beta_2$
  surface reconstruction.  (a) The locations  of  the adsorption sites
  are  indicated by stars.  (b)  Atomic configuration.   (c) Scheme of
  the transitions between  the A$_{i1}$ and A$_{i2}$ adsorption  sites
  and  the neighboring sites.  (d) Energy profile for barrier $T_{i1}$
  from the besides in-dimer adsorption site A$_4$ to A$_{i2}$.}
\label{b2-in}
\end{figure}

In this case we have two  As ad-dimers instead of  one.  We have found
one  adsorption site for the In  adatom for each  one of the ad-dimers
indicated in Figures  \ref{b2-ad} as A$_{a1}$  and A$_{a2}$.   We have
found that the In adsorbate does not lie in the vertical plane passing
through the As   dimers, but the  bonds  are tilted  towards the other
ad-dimer, that  is towards   the symmetry  plane  passing through  the
A$_{12}$ adsorption site.  Regarding the As in-dimer we have found two
distinct possibilities for In  adsorption that we indicate as A$_{i1}$
and A$_{i2}$   in   Figure \ref{b2-in}.  As  can   be  seen  from Fig.
\ref{b2-in}(b),  the In atom   bonds to  the  two  As of the  in-dimer
forming  a  tilted angle $\phi$   with respect   to  the vertical  $z$
direction.  Since the $(110)$  is a plane  of symmetry for the system,
the two configurations  for  the  adsorbate, corresponding  to  angles
$\phi$ and $-\phi$,  with respect to the  $z$ axis  are equivalent and
have the same probability to occur.

Figures \ref{b2-ad}(a) and   \ref{b2-in}(a) show the  location of  the
adsorption  sites for the  As ad-dimers and As in-dimer, respectively.
The  atomic configurations for  the   adsorbates are given in  Figures
\ref{b2-ad}(b) and   \ref{b2-in}(b), whereas  the transitions  to/from
neighboring adsorption sites are  reported in  Figures \ref{b2-ad}(c)
and \ref{b2-in}(c).  Finally Figures \ref{b2-ad}(d) and \ref{b2-in}(d)
plot  the energy barriers separating  these  adsorption sites from the
neighboring ones.

Tables \ref{tab3} and \ref{tab4} report  the  energies for the  stable
configurations and the corresponding saddle points.  Even in this case
we have  found very high binding energies  for  these adsorption sites
and   high   energy  barriers   (around  $0.7$   eV)  that  make these
configurations very stable once the adatom reaches them overcoming the
barriers.

\section{Discussion}
\begin{figure}[t]
\centering
\includegraphics*[width=8cm]{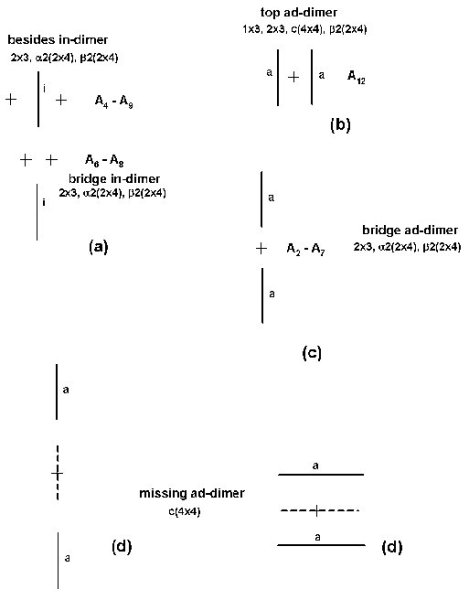}
\caption{Scheme of   the main adsorption  sites  for an  In  adatom on
  $(001)$  reconstructed III-V  surfaces:  (a)  besides   in-dimer and
  bridge  in-dimer locations, (b)   top ad-dimer location,  (c) bridge
  ad-dimer location, and (d) missing ad-dimer locations.}
\label{minima}
\end{figure}

In  order to evaluate the  effects of the strain  and  the role of the
different surface reconstructions on the  In adsorption properties  we
now compare  our  results  with  the  results of similar  calculations
reported  in the  literature.   To better evaluate  the differences we
report in Fig.  \ref{minima}   the more common adsorption  features as
extracted by the literature and the  present work.  In the same figure
we name  each specific  adsorption  site with respect  to its position
relative to the As in-dimers (i) and ad-dimers  (a), and indicate also
the reconstructions where these adsorption sites are typical.

\subsection{$\alpha_2$ vs $\beta_2$ reconstruction (this work)}
There are no big differences   between the adsorption features on  the
$\alpha_2$  and   on  the   $\beta_2$ surface    reconstructions.   In
particular, the adsorption energy  and location differences (see Table
\ref{tab5}) between corresponding  adsorption sites, are within  a few
hundreds of meV.  The greatest differences are  those in the region of
the missing  ad-dimer where the  structural difference between the two
surface reconstructions is larger. On the $\beta_2$ reconstruction the
positions and energies  of the adsorption  sites are modified owing to
the  symmetry  properties of the  reconstruction.   The most important
difference  is  the  presence of the   adsorption  sites  A$_{3'}$ and
A$_{12}$ on the  $\beta_2$ surface reconstruction,  that are absent on
the $\alpha_2$ surface reconstruction. On the contrary, the adsorption
sites  A$_{10}$ and A$_{11}$   of  $\alpha_2$ disappear  in  $\beta_2$
transforming into the top ad-dimer site A$_{12}$.

\subsection{1.75 ML InAs on GaAs $(001) \beta_2(2\times4)$ vs InAs $(001) \beta_2(2\times4)$}
In this subsection we compare  our results  for  In adsorption on  the
$\beta_2(2\times4)$ reconstruction of the InAs  WL on GaAs$(001)$ with
the   results     for  In adsorption    on   the   $\beta_2(2\times4)$
reconstruction of pure InAs$(001)$  of ref.   \cite{giappi1, giappi2}.
The main difference between the two cases is that  the InAs WL on GaAs
is subjected to a compressive strain  on the $(001)$ plane.  Thus, the
comparison allows  us to evaluate the  effect  of the WL  strain on In
adsorption. We principally refer to Fig.2 of ref.\cite{giappi1} and to
Table  \ref{tab6} of  the present paper,   where we have  reported the
energies of the minima of In on the InAs surface \cite{private}.

In    the case  of  pure   InAs\cite{giappi1}, only   six minima  were
calculated and their positions on  the surface are  almost the same as
those calculated by us. In  particular, only one  minimum B$_8$ in the
{\em bridge  in-dimer}   position (see Fig.    \ref{minima}) was found
halfway between the two equal minima  A$_6$ and A$_8$ calculated by us
for the  InAs/GaAs case. The two  minima A$_6$ and A$_8$ are separated
by a relatively   high barrier of $300$  meV  in our  calculation.  We
recall here  that for $\alpha_2$ we calculate  barriers  only $42$ and
$142$ meV high  between the same adsorption  sites.   (A$_6$ and A$_8$
are no more equivalent in $\alpha_2$).

In the case of  pure InAs, the {\em besides  in-dimer} B$_4$ and B$_9$
show  the strongest binding.  In the  case of  the  InAs  WL on  GaAs,
instead, the more strongly bonded sites are located at the {\em bridge
  in-dimer}  positions A$_6$ and   A$_8$.  From Tables  \ref{tab3} and
\ref{tab6}   we  see  that  the energy  difference  between  the sites
($4$-$9$) and ($6$-$8$) is larger  for the InAs  WL on GaAs case  (240
meV vs 61 meV).  It is reasonable to think  that this difference is to
be  ascribed to the strain originated  by the lattice mismatch between
InAs and GaAs.   Also  the absence   of the  shallow adsorption  sites
corresponding to A$_1$, A$_5$,  A$_3$ and A$_{10}$ could be originated
by the absence of the strain.

Nevertheless,  it is  possible that  these differences are  originated
also by the different grid resolution used for calculating the PES, or
by   the interpolation algorithms  that    sometimes are not able   to
identify correctly the minima or by the difference between GGA used in
ref.  \cite{giappi1} and LDA used in this work.

\subsection{In on InAs/GaAs $\beta_2(2\times4)$ vs Ga on GaAs $\beta_2(2\times4)$}
At this stage  of   the discussion  we  compare  our results  for  the
$\beta_2$  reconstruction  with   the   results   obtained   in   ref.
\cite{ruggerone},   where   the   PES   of  a    Ga  adatom    on  the
$\beta_2(2\times4)$ surface reconstruction of GaAs$(001)$ was studied.
This comparison  evidences the different     behavior of In  and    Ga
adsorbates, as opposed to the WL composition.

In  the case of Ga   on GaAs$(001)$ only   three minima were  reported
(without   considering the   additional minima  where   Ga  breaks the
dimers).  Two minima are at {\em  bridge ad-dimer} positions while the
third one is  at a {\em bridge  in-dimer}  position.  These adsorption
sites  correspond to the  A$_2$,  A$_7$ and A$_6$-A$_8$  minima on the
InAs WL and  to B$_2$, B$_7$ and  B$_8$ minima of  the In on InAs(001)
case   \cite{giappi1}.  These  minima    are quite stable. Notice  the
missing of  the  A$_{12}$ adsorption  site on  the {\em top  ad-dimer}
location,  which is   present in the   case of   the In adsorbate   on
$\beta_2$ reconstructed   InAs covered  surfaces. Also the  adsorption
sites in  the {\em  besides in-dimer} locations   are missing in  this
case.

\subsection{In adatom on InGaAs surfaces}
In   this section  we  compare with   the  results of   ref.
\cite{penevpap,  penevc4x4} where InGaAs reconstructions  of the WL on
GaAs   were   considered.  The  PESs    of the    In   adatom  on: (i)
In$_{2/3}$Ga$_{1/3}$As$(001)$   $(2\times3)$  (corresponding to  a  In
coverage   $\theta=0.66$),       (ii)    In$_{2/3}$Ga$_{1/3}$As$(001)$
$(1\times3)$  (corresponding to  $\theta=0.66$), and (iii) GaAs$(001)$
$c(4\times4)$  (corresponding  to $\theta=0$)  surface reconstructions
were studied.    From all these   calculations we  can see   that  one
adsorption site is always located in a {\em top ad-dimer} position, in
the same configuration of the A$_{12}$ adsorption site on $\beta_2$ of
this work.  The depth of these  adsorption sites defined by the height
of the lowest barrier confining   the adsorbate are: $140$ meV,   $60$
meV,   $100$  meV,    $385$  meV  and $135$    meV  for  $(2\times3)$,
$(1\times3)$, $c(4\times4)$, InAs  $\beta_2(2\times4)$ and InAs WL  on
GaAs $\beta_2(2\times4)$, respectively.  We  can see that these minima
are not strongly bonded, apart from the case of In on pure InAs$(001)\
\beta_2$ reconstructed surface.

The deepest minima are located at the {\em besides in-dimer} location,
both in the case of the $(2\times3)$ reconstruction of \cite{penevpap}
and of the  reconstructions studied in this work.   In the case of the
$c(4\times4)$ reconstruction, stable locations are at the {\em missing
  ad-dimer} sites (see fig.  \ref{minima}).  In these adsorption sites
the  In adatom  binds to  four As  atoms below.  In   the case  of the
$(1\times3)$ reconstruction the   stable  location sees the In   atoms
bonded to the cations layer below.

\section{Conclusions}
We have  carried   out a first-principles   study  of  the  adsorption
properties of an In adatom deposited on a InAs wetting
layer on GaAs$(001)$ reconstructed  $\alpha_2 (2\times4)$ or $\beta_2
(2\times4)$.  We  have first studied the   equilibrium geometry of the
surfaces and the relative  phase stability with  respect to the growth
conditions. We  have  found that,  in the case  of   the InAs  WL  the
interlayer distances  are  systematically  higher than those   of pure
GaAs, owing to the compressive   strain. Moreover, we have found  that
the $\alpha_2$  reconstruction  of the InAs  WL  is  favored over  the
$\beta_2$ reconstruction, for a wider range of growth conditions, when
compared to the bare GaAs$(001)$ surface.

We have analyzed  the PES for  an In adatom determining the adsorption
sites  and   the  saddle points for  both    $\alpha_2$  and $\beta_2$
reconstructions.   We   have found $11$  minima  in  the PES  for both
reconstructions.  Most of  them  occur in  the same positions  in both
cases.    Two  minima are instead    specific  of  the $\alpha_2$  and
$\beta_2$ reconstructions,  due to  the presence  on  $\beta_2$  of an
additional As ad-dimer.   For   the adsorption sites present  on  both
reconstructions,  we have studied  the change of the adsorption energy
and location  caused  by  the presence  or  absence of  the  second As
ad-dimer.   We have   identified two  low-energy  trenches  along  the
$[\bar110]$  direction,  alongside  the   in-dimer   chain, for   both
$\alpha_2$ and $\beta_2$.

The comparison to  analogous  studies reported in the   literature has
revealed some   general features for   the In adsorption  on different
surface  reconstructions.  In  particular,  we have found  that stable
adsorption  sites are always located: (i)  besides the in-dimers, (ii)
at  the bridge  positions  between in-dimers,  (iii) between ad-dimers
(only for $\beta_2$ which has two  adjacent As ad-dimers), and (iv) at
the bridge between ad-dimer  positions.  We have identified also other
shallower adsorption sites which are  more reconstruction specific and
are related to $\alpha_2$ having a lower symmetry than $\beta_2$.

\begin{acknowledgments}
  The authors would like to thank A.  Ishii for  providing us with the
  data about In adsorption on InAs \cite{giappi1}.  We acknowledge the
  support   of the MIUR     PRIN-2005, Italy.  The  calculations  were
  performed at  CINECA-Bologna under  the grant   ``Iniziativa Calcolo
  Parallelo del CNR-INFM'', and at LabCsai in Modena.
\end{acknowledgments}

\appendix
\section{Energies of minima and saddle points}
In  this appendix we  report all  the numerical  values calculated for
minima and saddle point energies.

\begin{table}
\centering
\begin{tabular}{lc}
\hline
site & energy (meV) \\
\hline
\hline
A$_1$    &  220 \\
A$_2$    &  387 \\
A$_3$    &  312 \\
A$_4$    &  49  \\
A$_5$    &  344 \\
A$_6$    &  99  \\
A$_7$    &  559 \\
A$_8$    &  0	\\
A$_9$    &  112 \\
A$_{10}$ &  868 \\
A$_{11}$ &  498 \\
\hline
A$_i$    & -100 \\
A$_a$    &  239 \\
\hline
\end{tabular}
\caption{Adsorption energies for the In adatom in the different adsorption sites of the  $\alpha_2$ surface reconstruction, relative to the A$_8$ energy, labeled as A$_n$. The labels A$_i$ and A$_a$ refer to the configurations of the adsorbate into the in-dimer and the ad-dimer, respectively.}
\label{tab1}
\end{table}

\begin{table}
\centering
\begin{tabular}{lc}
\hline
saddle & energy (meV) \\
\hline
\hline
T$_{2,9}$   & 895 \\
T$_{2,7}$   & 676 \\
T$_{3,7}$   & 635 \\
T$_{3,4}$   & 456 \\
T$_{4,9}$   & 498 \\
T$_{4,5}$   & 453 \\
T$_{1,10}$  & 893 \\
T$_{6,8}$   & 142 \\
T$_{8,9}$   & 428 \\
T$_{1,9}$   & 272 \\
T$_{1,8}$   & 418 \\
T$_{5,11}$  & 554 \\
T$_{10,11}$ & 929 \\
T$_{7,11}$  & 743 \\
T$_{11,7}$  & 750 \\
\hline
T$_{i1}$ & 613   \\
T$_{i2}$ & 493   \\
\hline
T$_{a1}$ &  893  \\
T$_{a2}$ &  926  \\
T$_{a3}$ &  1117 \\

\hline
\end{tabular}
\caption{Energies of the saddle points in the PES of the  $\alpha_2$ surface reconstruction, relative to the A$_8$ adsorption energy. The saddle points connecting the minima $A_n$ and $A_m$ are labeled by T$_{n,m}$, while the labels T$_{in}$ (T$_{an}$) indicate the saddle points between the minima into the in-dimer (ad-dimer) and the neighboring sites, as shown in Fig. \ref{a2-in} (Fig. \ref{a2-ad})}
\label{tab2}
\end{table}

\begin{table}
\centering
\begin{tabular}{lc}
\hline
site & energy (meV) \\
\hline
\hline
A$_1$, A$_5$    &  216 \\
A$_2$, A$_7$    &  506 \\
A$_3$, A$_{3'}$ &  323 \\
A$_4$, A$_9$    &  240 \\
A$_6$, A$_8$    &  0   \\
A$_{12}$        &  588 \\
\hline
A$_i$           & -330 \\
A$_a$           &  158 \\
\hline
\end{tabular}
\caption{Adsorption energies  for  the  In adatom  in  the different adsorption sites of the $\beta_2$ surface reconstruction, relative to the A$_8$ energy, labeled as A$_n$. The labels A$_i$ and A$_a$ refer to the configuration of the adsorbate into the in-dimer and the ad-dimer, respectively.}
\label{tab3}
\end{table}

\begin{table}
\centering
\begin{tabular}{lrr}
\hline
site & \quad $\Delta E$ (meV) & \quad ${\Delta r}/{a_0}$ \\
\hline
\hline
A$_1$ & -134 & 0.022 \\
A$_2$ &   -7 & 0.069 \\
A$_3$ & -121 & 0.330 \\
A$_4$ &   66 & 0.168 \\
A$_5$ & -259 & 0.350 \\
A$_6$ & -224 & 0.100 \\
A$_7$ & -179 & 0.588 \\
A$_8$ & -125 & 0.018 \\
A$_9$ &    4 & 0.031 \\
\hline
\end{tabular}
\caption{Adsorption energy differences and position displacements between the In adsorbate for the corresponding adsorption sites of the $\beta_2$ and the $\alpha_2$ reconstructions.}
\label{tab5}
\end{table}

\begin{table}
\centering
\begin{tabular}{lc}
\hline
saddle & energy (meV) \\
\hline
\hline
T$_{4,5}$   &  392  \\
T$_{5,6}$   &  424  \\
T$_{1,12}$  &  976  \\
T$_{12,2}$  &  723  \\
T$_{2,11}$  &  727  \\
T$_{2,3'}$  &  974  \\
T$_{2,7}$   &  589  \\
T$_{6,3}$   &  577  \\
T$_{4,3}$   &  424  \\
T$_{4,9}$   &  701  \\
T$_{6,8}$   &  296  \\
\hline
T$_{i1}$ &  589  \\
T$_{i2}$ &   94  \\
\hline
T$_{a1}$ &  760  \\
T$_{a2}$ &  610  \\
T$_{a3}$ &  1024 \\
\hline
\end{tabular}
\caption{Energies of the saddle points in the PES of the  $\beta_2$ surface reconstruction, relative to the A$_8$ ($\beta_2$) adsorption energy. The saddle points connecting the minima $A_n$ and $A_m$  are labeled by T$_{n,m}$, while the labels T$_{in}$ (T$_{an}$) refer to the saddle points between the minima into the in-dimer (ad-dimer) and the neighboring sites, as shown in Fig. \ref{b2-in} (Fig. \ref{b2-ad}).}
\label{tab4}
\end{table}

\begin{table}
\centering
\begin{tabular}{lc}
\hline
site & energy (meV)  \\
\hline
\hline
B$_2$, B$_7$   & 302 \\
B$_4$, B$_9$   &   0 \\
B$_8$          &  61 \\
B$_{11}$       & 163 \\
\hline
\end{tabular}
\caption{Adsorption energies (calculated with respect to B$_4$ energy) for  the  In adatom in  the different adsorption sites on the InAs $\beta_2$ surface reconstruction, labeled as B$_n$ (see ref. \cite{giappi1, private}). The positions of the minima B$_n$ corresponds to the minima A$_n$ in fig. \ref{pb}.}
\label{tab6}
\end{table}

\end{document}